\documentclass[doublecol]{epl2} 
\usepackage{amsmath}
\usepackage{graphicx}
\usepackage{subfigure}
\usepackage{ amssymb }

\title{Discrete-time thermodynamic uncertainty relation}

\author{Karel Proesmans \and Christian Van den Broeck}

\institute{                    
Hasselt University, B-3590 Diepenbeek, Belgium
}

\abstract{We generalize the thermodynamic uncertainty relation, providing an entropic upper bound for average fluxes in time-continuous steady-state systems (Gingrich et al., Phys. Rev. Lett. 116, 120601 (2016)), to time-discrete Markov chains and to systems under time-symmetric, periodic driving. }

\begin{document}

\maketitle

\section{Introduction}
There are several ways to characterize a  system in nonequilibrium. Such a system breaks time-reversal invariance. It does not obey detailed balance. It dissipates. It possesses non-zero fluxes.
Very recently, a surprising inequality was discovered that links  these concepts \cite{barato2015thermodynamic,pietzonka2016universal2,gingrich2016dissipation}.
The inequality states that the average of a thermodynamic-like flux, $\overline{j}$, (such as work, heat or particle flux) is bounded by the variance of its fluctuations $\overline{\delta j^2}$ and the total rate of entropy production in the system, $\dot{S}_i$, in the following way:
\begin{equation}
    \frac{\overline{j}^2}{\overline{\delta j^2}}\leq \frac{\dot{S}_i}{2k_B}.\label{turor}
\end{equation}
$k_B$ is Boltzmann's constant. The great interest of this  result is that it is valid "arbitrary far from equilibrium". Furthermore, its usefulness has, in its short existence, been illustrated in different contexts including  molecular motors \cite{pietzonka2016universal}, first-passage problems \cite{garrahan2017simple,gingrich2017fundamental}, heat engines \cite{pietzonka2017universal}, self-assembly \cite{nguyen2016design}, information theory \cite{horowitz2017information}, and biochemical oscillations \cite{barato2017persistence}.
Originally, the above relation was obtained as a long-time result for systems with a finite state space \cite{gingrich2016dissipation}. More recently, it was shown to hold in finite-time \cite{pietzonka2017finite,maes2017frenetic,horowitz2017proof} and for diffusive systems \cite{polettini2016tightening,hyeon2017physical,gingrich2017inferring}. 
 It is, however, not valid for time-discrete Markov chains \cite{shiraishi2017finite} or systems with explicit time-dependent driving \cite{barato2016cost,rotskoff2017mapping,1751-8121-50-35-355001}. This raises the 
  question as to whether there exists a generalization  that covers these situations.

In this letter, we (partially) answer this question by deriving the following generalized uncertainty relation:
\begin{equation}
    \frac{\overline{j}^2}{\overline{\delta j^2}}\leq \frac{1}{2\Delta t}\left(e^{\Delta_i S/{ k_B}}-1\right).\label{turne}
\end{equation}
It is valid for Markov chains and for periodically driven systems under the extra assumption of time-symmetric driving (i.e.,~the driving is invariant under time-reversal). $\Delta t$ is the duration of one Markov step or period of the driving and $\Delta_i S$ is the associated entropy production. The uncertainty relation  Eq~(\ref{turor}) is recovered in the continuous time limit, $\Delta t\rightarrow 0$ with $\Delta_i S/\Delta t\rightarrow \dot{S}_i$.

The outline of this letter is  as follows. Our derivation in built on the large deviation properties of empirical distribution for Markov chains, which are reviewed in section $2$. In section $3$, we derive the generalized uncertainty relation, Eq.~(\ref{turne}), and illustrate the inequality for a random walker and a two-level system in section $4$. We conclude with a short discussion.

\section{Large Deviation theory}
Consider a time-homogeneous  irreducible Markov chain, characterized by the  (time-independent, pairwise) probability ${\bf p}=\{p_{kl}\}$,  with $p_{kl}$ to probability to be at a given time in state $k$ and go to state $l$ in a single time step $\Delta t$.  The (pair) empirical distribution ${\bf q}=\{ q_{kl}\}$ is defined as the observed probability for the pair states  observed in a finite run., i.e.,  $q_{kl}$ is equal to the fraction of  pairs  $k$ followed by $l$, that is observed in a  run of  $N=t/\Delta t$ steps. In the long time limit, $N\rightarrow \infty$, $q_{kl}$ will converge to $p_{kl}$.  According to the theory of  large deviations, the asymptotic convergence is such that any other empirical density becomes exponentially unlikely, i.e.,
\begin{equation}
P_t({\bf q})=\exp\left(-t\mathcal{I}\left({\bf q}\right)+o(t)\right)\label{LDF1},
\end{equation}
or
\begin{equation}
\mathcal{I}\left({\bf q}\right)=-\lim_{t\rightarrow \infty}\frac{1}{t}\ln P\left({\bf q}\right).
\end{equation}
$\mathcal{I}\left({\bf q}\right)$ is called the large deviation function associated with the empirical density. It satisfies
\begin{equation}
\mathcal{I}\left({\bf q}\right)\geq 0,\quad \mathcal{I}\left({\bf q}\right)=0\Leftrightarrow q_{kl}=p_{kl},\quad \forall k,l.\label{ldfprop}
\end{equation}
The explicit expression of this large deviation function is known \cite{den2008large,touchette_large_2009}:
\begin{equation}
\mathcal{I}\left({\bf q}\right)=\frac{1}{\Delta t}\left(\sum_{k,l}q_{kl}\ln\left(\frac{q_{kl}}{p_{kl}}\right)-\sum_{k}q_{k}\ln\left(\frac{q_k}{p_k}\right)\right),\label{edldf}
\end{equation}
with $q_k=\sum_l q_{kl}$ and $p_k=\sum_l p_{kl}$. 

We are interested  in the large deviation properties of a "reduced" quantity, namely a generic thermodynamic flux $j$. It is 
 a linear combination of the net empirical fluxes between any two states $k$ and $l$ \cite{schnakenberg1976network}:
\begin{equation}
j=
\sum_{k,l}\mathcal{F}_{kl}q_{kl},
\end{equation}
with $\mathcal{F}$ an antisymmetric matrix, $\mathcal{F}_{kl}=-\mathcal{F}_{lk}$. The large deviation function of $j$ is defined as
\begin{equation}
\mathcal{J}(j)=-\lim_{t\rightarrow \infty}\frac{1}{t}\ln P_t(j).
\end{equation}
This function is again non-negative, and will only be zero for $j$ equal to its "true" average $j=\overline{j}$:
\begin{equation}\label{jbar}
\overline{j}=\sum_{k,l}\mathcal{F}_{kl}p_{kl}.
\end{equation} 
It can be obtained from the  large deviation function for the empirical density via the so-called contraction principle \cite{touchette_large_2009}:
\begin{equation}
\mathcal{J}(j)=\min_{\{q_{kl},\sum_{k,l}\mathcal{F}_{kl}q_{kl}=j\}}\mathcal{I}\left({\bf q}\right),\label{contr}
\end{equation}
{where $q_{kl}$ should also satisfy the properties of an empirical density, i.e., $\sum_{k,l}q_{kl}=1$,  $q_{kl}\geq 0$ and $\sum_k q_{kl}=\sum_l q_{kl}$.}
We finally mention that   $\mathcal{J}(j)$ (typically) has a parabolic minimum around $\mathcal{J}(\overline{j})=0$, with second derivative related to the variance of $j$, $\overline{\delta j^2}$, as follows:
\begin{equation}
\mathcal{J}''(\overline{j})=\frac{1}{\overline{\delta j^2}}.\label{vardef}
\end{equation}

\section{Thermodynamic uncertainty relation}
To derive the generalized thermodynamic uncertainty relation, Eq.~(\ref{turne}), we start from the contraction principle, Eq.~(\ref{contr}). The implied constrained optimization is difficult to perform. Instead, an upper bound can  be obtained using the following trial empirical density:
\begin{equation}
q^j_{kl}=p_{kl}+\frac{j-\overline{j}}{\overline{j}}\left(p_{kl}-\frac{p_{kl}p_{lk}}{\mathcal{N}\left(p_{kl}+p_{lk}\right)}\right),
\end{equation}
with
\begin{equation}
\mathcal{N}=\sum_{k,l}\frac{p_{kl}p_{lk}}{p_{kl}+p_{lk}}.
\end{equation}
One can easily verify that this density is normalized. More importantly, the antisymmetry of $\mathcal{F}$ together with Eq.~(\ref{jbar}) implies that the constraint $\sum_{k,l}\mathcal{F}_{kl}q^j_{kl}=j$ is automatically satisfied. 
We thus conclude from Eq.~(\ref{contr}):
\begin{equation}
\mathcal{J}(j)\leq I\left(\{q^{j}_{kl}\}\right)\leq \frac{1}{\Delta t}\sum_{k,l}q^{j}_{kl}\ln\left(\frac{q^{j}_{kl}}{p_{kl}}\right),
\end{equation}
where we used the fact that $\sum_kq^{j}_k\ln (q^{j}_k/p_k)\geq 0$.
Note that the left and right hand side of the above relation, as well as their first derivatives with respect to $j$, are all equal to zero for $j=\bar{j}$. An expansion up to second order around this value together with Eq.~(\ref{vardef}) thus leads to the following inequality:
\begin{equation}
\frac{1}{\overline{\delta j^2}}\leq \frac{1}{\overline{j}^2\Delta t}\left(\frac{1}{2\mathcal{N}}-1\right).\label{nb}
\end{equation}

Next we introduce the entropy production for a step in the Markov chain:
\begin{equation}
\Delta_i S=k_B\sum_{k,l} p_{kl}\ln\left(\frac{p_{kl}}{p_{lk}}\right).\label{s1}
\end{equation}
While this definition appears to be in agreement with stochastic thermodynamics \cite{schnakenberg1976network,jiu1984stability,seifert_stochastic_2012,van_den_broeck_ensemble_2014},
we stress that a proper thermodynamic interpretation requires additional input about the physics of the system, for example about the energies of the different states, as well as properties of the transition matrix such as local detailed balance. For this reason, the  inequality derived below is of statistical
origin, resting only on generic properties of Markov chains.
To make now the connection between the entropy production and the flux, we refer to the appendix for the derivation of the following inequality:
\begin{equation}
\frac{1}{\mathcal{N}}\leq{e^{\Delta_i S/{k_B}}+1}.\label{sb}
\end{equation}
The thermodynamic uncertainty relation for Markov chains follows by combination with Eq.~(\ref{nb}).

The above derivation can be adapted to periodically driven systems  with time-symmetric driving. $\Delta t$ now plays the role of one period. The trajectory of the system over one such period is denoted by $\Gamma$, where $\Gamma(t)$ denotes the state of the system at time $t$. Associated with every path, there is a time-inverted path defined by $\tilde{\Gamma}(t)=\Gamma(\Delta t-t)$, so that $\tilde{\tilde{\Gamma}}(t)=\Gamma(t)$. The uncertainty relation, Eq.~(\ref{turne}) is valid for fluxes of the form:
 \begin{equation}j=
 \sum_{\Gamma}\mathcal{F}_\Gamma q_\Gamma,\end{equation}
where $q_{\Gamma}$ denotes the empirical distribution to observe the trajectory ${\Gamma}$, i.e. the fraction of periods in which it is observed in a run of $N=t/\Delta t$ periods,  and $\mathcal{F}_{\tilde{\Gamma}}=-\mathcal{F}_\Gamma$ is antisymmetric with respect to time-reversal. In the appendix, we {provide a handwaving derivation} for the large deviation associated with $q_{\Gamma}$:
\begin{equation}
I(\{q_{\Gamma}\})=\frac{1}{\Delta t}\left(\sum_{\Gamma}q_\Gamma\ln\left(\frac{q_\Gamma}{p_\Gamma}\right)-\sum_k q_k\ln\left(\frac{q_k}{p_k}\right)\right),\label{edldf2}
\end{equation}
where $p_\Gamma$ is the probability that the state of the system during one period is described by $\Gamma$, and $q_k$ and $p_k$ are the empirical density and probability for the system to be in state $k$ at the beginning of a cycle. In agreement with  stochastic thermodynamics for periodically perturbed systems and using the assumption of time-symmetric driving, we define the entropy production over one cycle as \cite{kawai2007dissipation,gomez2008footprints}:
\begin{equation}
\Delta_i S=k_B\sum_\Gamma p_\Gamma\ln\left(\frac{p_\Gamma}{p_{\tilde{\Gamma}}}\right).\label{s2}
\end{equation}
The inequality Eq.~(\ref{turne}) follows by observing that Eqs.~(\ref{edldf2}) and (\ref{s2}) are identical to Eqs.~(\ref{edldf}) and (\ref{s1}) upon replacement   of $q_{kl}$ and $q_{lk}$ by $q_\Gamma$ and $q_{\tilde{\Gamma}}$

\section{Examples}
\begin{figure}

\subfigure[]{\includegraphics[width=8cm]{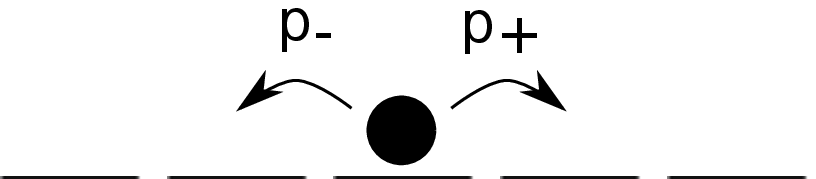}\label{opsrw}}\\
\subfigure[]{\includegraphics[width=8cm]{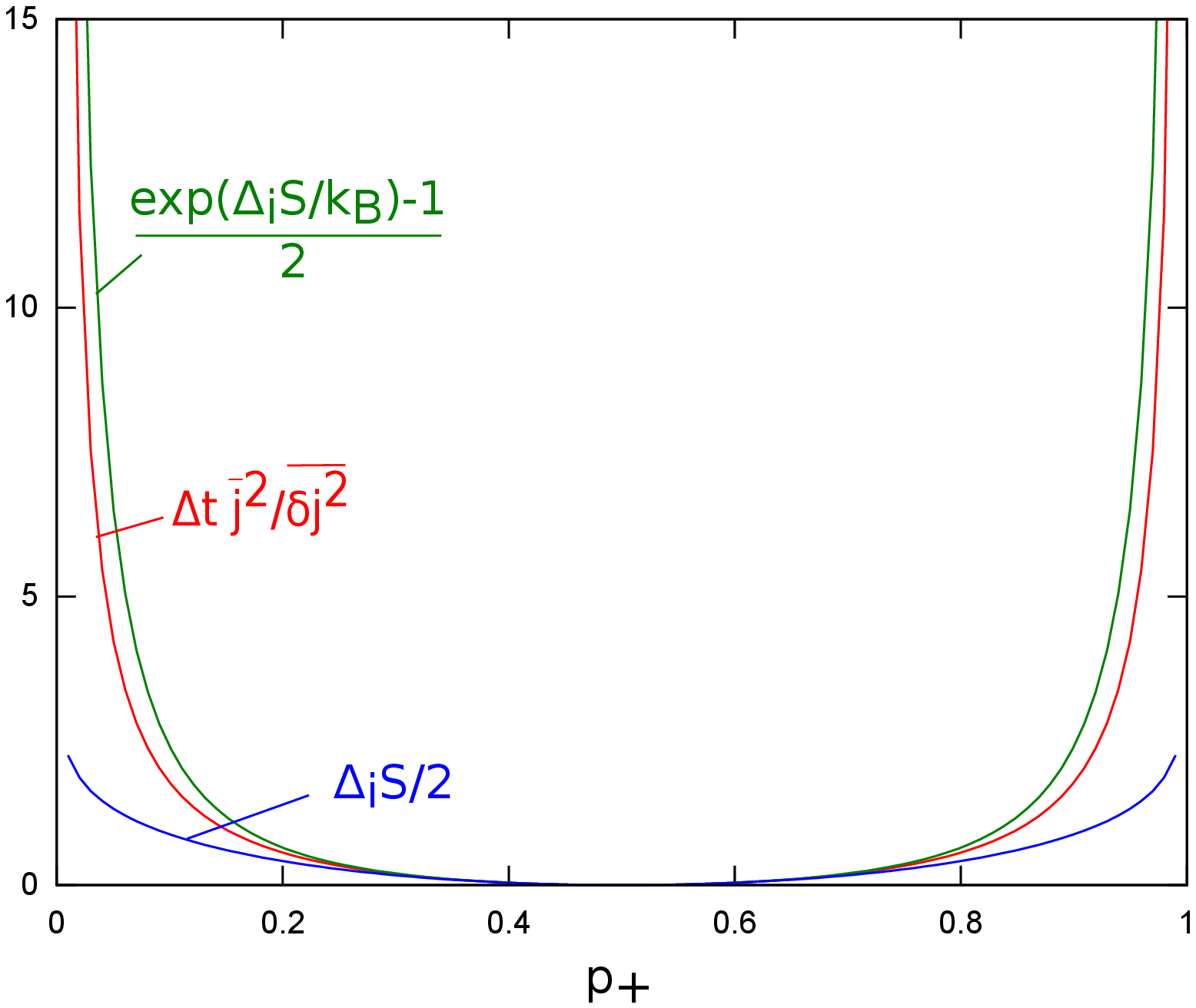}\label{turrw}}
\caption{a) Schematic representation of a time-discrete random walker. b) The thermodynamic uncertainty relation Eq.~(\ref{turne}) is valid for all values of $p_+$, while Eq.~(\ref{turor}) is not.}
\end{figure}
We illustrate the thermodynamic uncertainty relation on two simple examples. We first consider a biased discrete-time random walk, cf.~Fig.~\ref{opsrw}. Let $p_+$ and $p_-=1-p_+$ be the probability per time step  to go to the right and left, respectively. We focus on the stochastic rate $j$ for a particle to move to the right, being the net number of jumps to the right divided by the time (total number of jumps times $\Delta t$)). Its first two (central) moments are given by:
\begin{equation}
\overline{j}=\frac{p_{+}-p_-}{\Delta t},\qquad \overline{\delta j^2}=\frac{4p_-p_+}{\Delta t},
\end{equation}
while the entropy production per time step reads:
\begin{equation}
\Delta_i S=k_B(p_+-p_-)\ln\frac{p_+}{p_-}.
\end{equation}
The thermodynamic uncertainty relation, Eq.~(\ref{turne}), is reproduced:
\begin{multline}
    \frac{\overline{j}^2 \Delta t}{\overline{\delta j^2}}=\frac{(1-2p_+)^2}{4p_+(1-p_+)}\leq \frac{\left(\frac{p_+}{1-p_+}\right)^{2p_+-1}-1}{2}\\= \frac{1}{2}\left(e^{\Delta_i S/{ k_B}}-1\right).
\end{multline}
cf.~Fig.~\ref{turrw}. Note that the bound becomes tight in the limit of an unbiased walker, and remains qualitatively correct (same type of divergence, but with extra prefactor $2$)  in the limit of a one-sided walker ($p_+ \rightarrow 0$ or $1$).

\begin{figure}

\subfigure[]{\includegraphics[width=8cm]{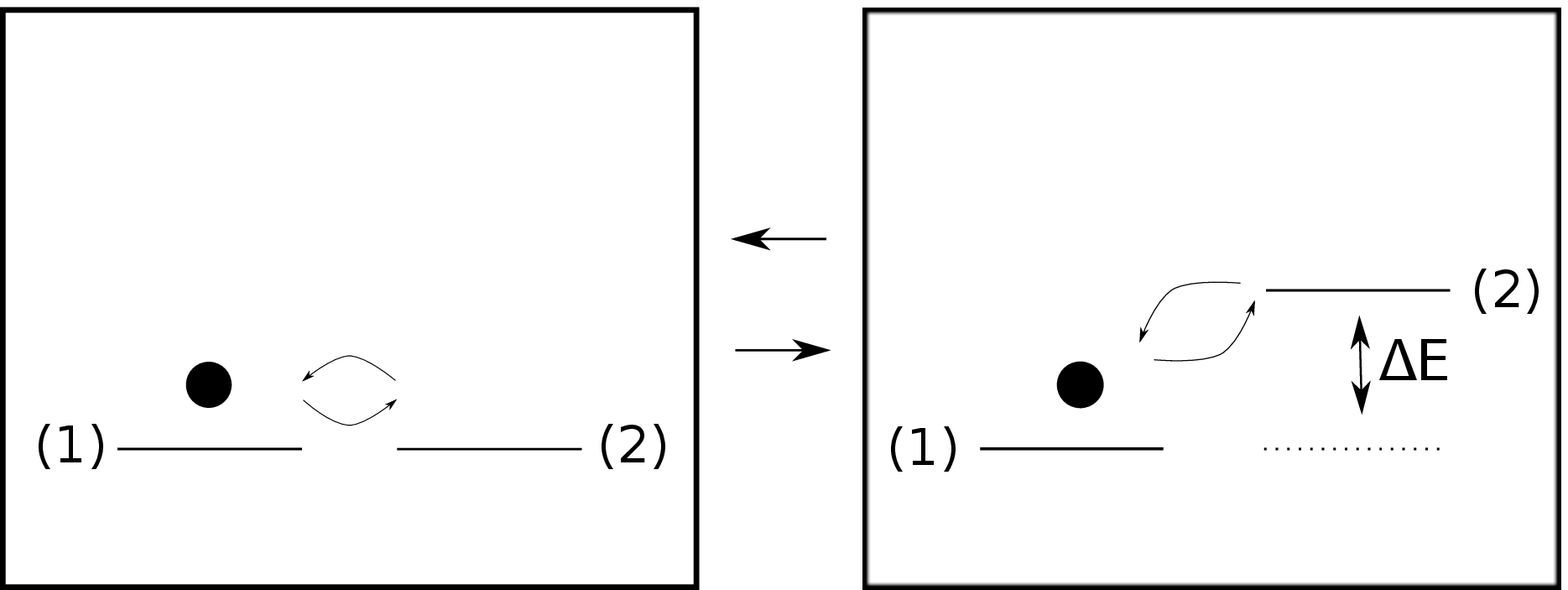}\label{ops1}}\\
\subfigure[]{\includegraphics[width=8cm]{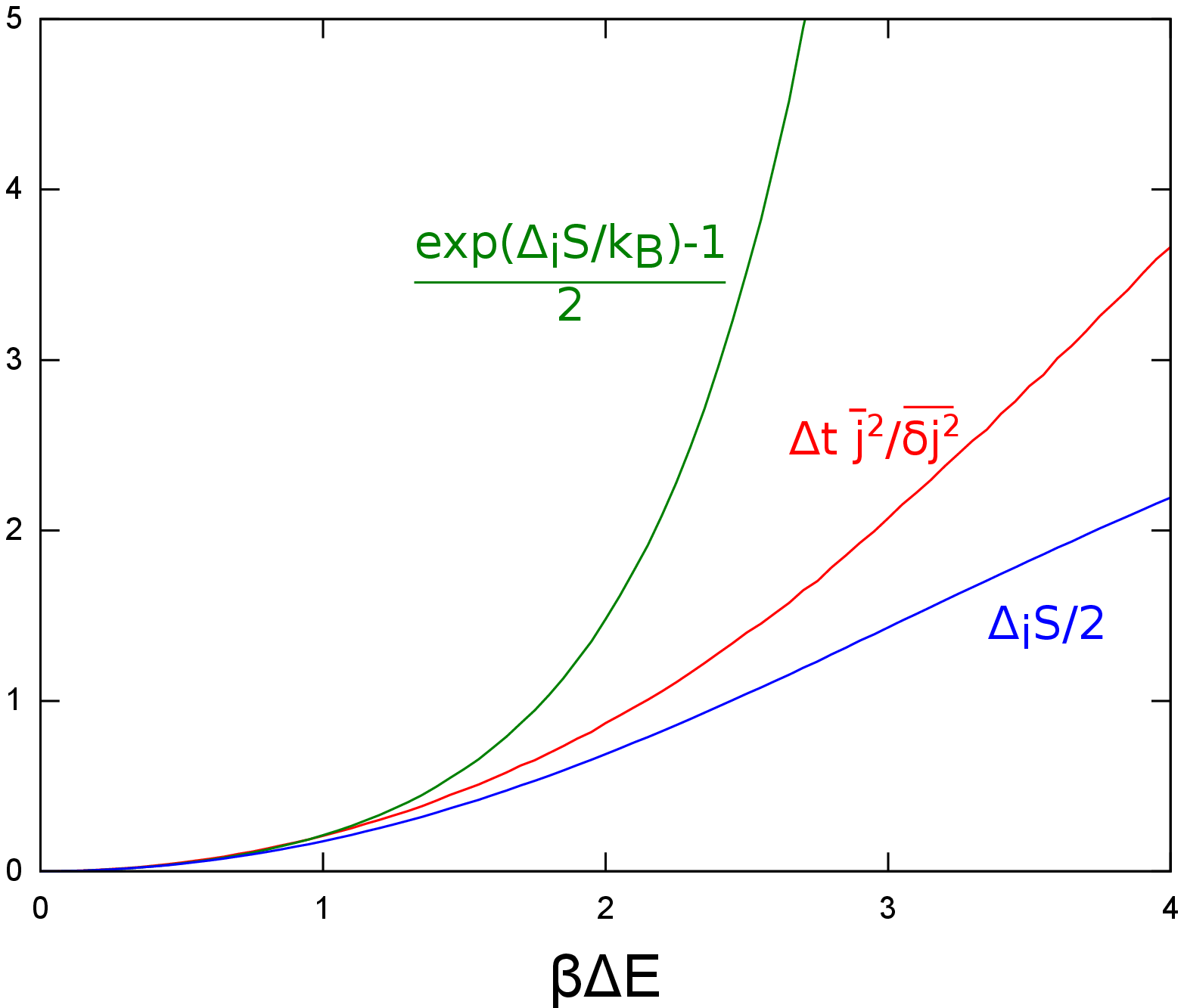}\label{tur2}}\\
\subfigure[]{\includegraphics[width=8cm]{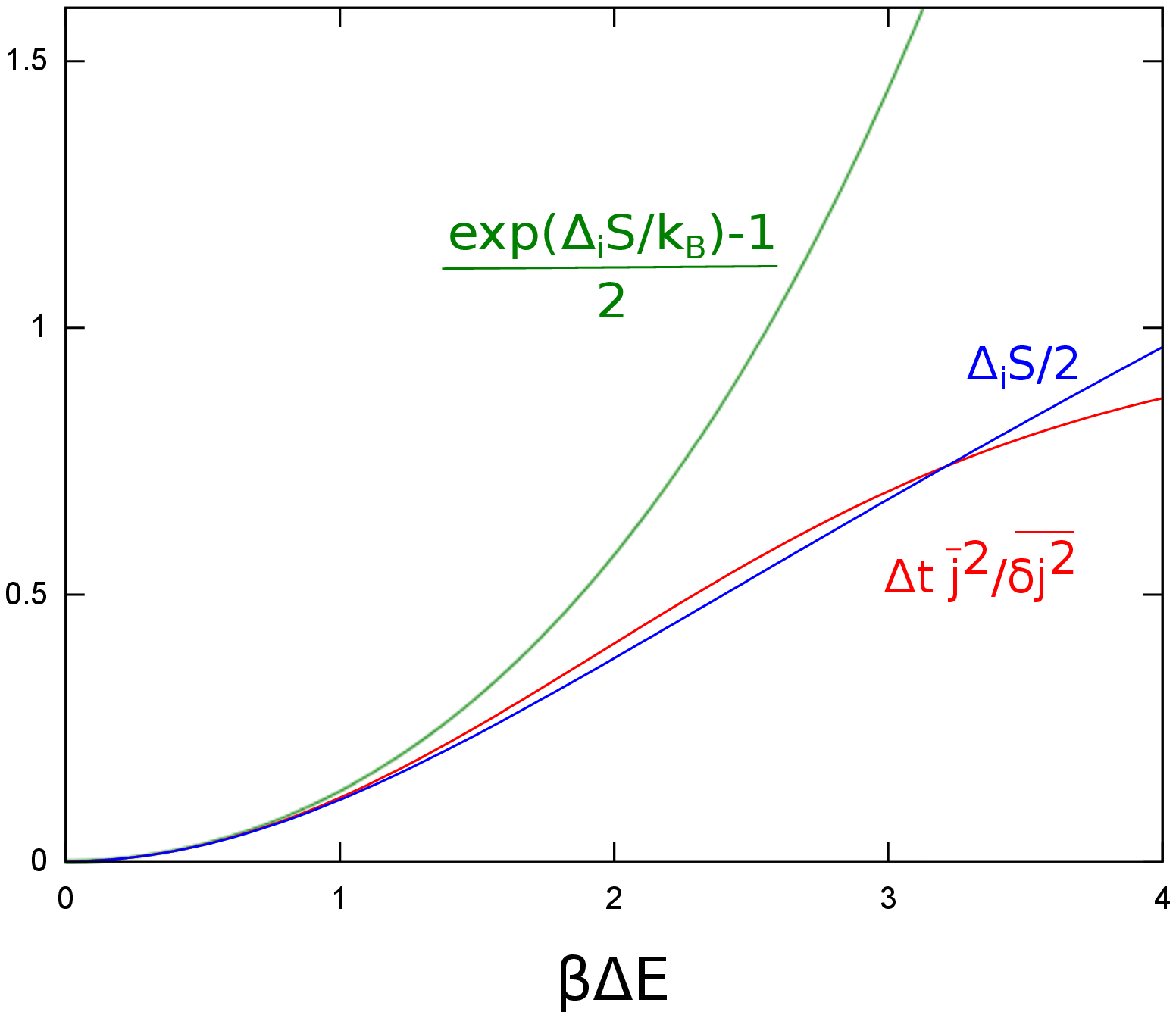}\label{tur1}}
\caption{a) Schematic representation of a two-level system with time-symmetric driving. {b) Numerical results for the thermodynamic uncertainty relation for cosine driving with rate constant, $K=2/\Delta t$. Eq.~(\ref{turne}) is valid, while Eq.~(\ref{turor}) is not. c) The thermodynamic uncertainty relation for the analytically solvable two-level system. Eq.~(\ref{turne}) is again valid for all values of $\beta\Delta E$, while Eq.~(\ref{turor}) is violated for $ \beta\Delta E\lesssim 3$.}}
\end{figure}
Next, we test the bound on a two-state periodically driven system, in contact with a thermal reservoir at temperature $T$.  We have in mind a quantum dot, in which one of two active energy levels is modulated by an external field. {The particle can jump from state $1$ to state $2$ with rate
\begin{equation}
W_{21}(t)=K\exp\left(\beta \left(E_1-E_2\right)\right),
\end{equation}
and vice versa. Here, $K$ is a rate constant, $\beta=1/(k_BT)$ and $E_i$, $i=1,2$, are the energies associated with the states. 
We consider two protocols with time-symmetric driving of the energy level $2$ (see Fig.~\ref{ops1}).  Our focus will be on the heat flux into the system: for every transition from state $i$ to $j$ at a time $t$, an amount of heat equal to $E_j(t)-E_i(t)$ is extracted from the heat bath.
First, we consider a cosine driving of level $2$: 
\begin{equation}
    E_1(t)=0,\qquad E_2(t)=\Delta E\cos\left(\frac{2\pi t}{\mathcal{T}}\right).
\end{equation}
We test the thermodynamic uncertainty relation for the heat flux $j$ into the system via numerical simulations \cite{holubec2011attempt}, cf. Fig.~\ref{tur2}:  the thermodynamic relation, Eq.~(\ref{turne}) is indeed verified, while Eq.~(\ref{turor}) is not.

Secondly, we consider the piece-wise constant modulation of level $2$, and derive exact results in the slow modulation limit.
Both levels start with the same energy $E_1=E_2=0$.  Next, the energy of level $2$ is lifted to $E_2=\Delta E$. We assume that the relaxation rate is fast  (or modulation slow) so that the system relaxes to the equilibrium distribution:
\begin{equation}
p_2=\frac{e^{-\beta \Delta E}}{e^{-\beta \Delta E}+1}=1-p_1.\label{pex}
\end{equation}}
Following this relaxation, the energy of level $2$ is again lowered to $E_2=0$, and the system again relaxes to the corresponding equilibrium state $p_1=p_2=1/2$.  {We again focus on the heat flux ${j}$} produced during this cycle. The average heat flux per cycle is:
\begin{equation}
\overline{j}=\left(\frac{1}{2}-p_2\right)\frac{\Delta E}{{\Delta t}}=\frac{1-e^{-\beta\Delta E}}{2\left(e^{-\beta\Delta E}+1\right)}\frac{\Delta E}{\Delta t}.
\end{equation}
By a similar argument, one can derive the variance:
\begin{equation}
\overline{\delta j^2}=\overline{j^2}-\overline{j}^2=\frac{\left(e^{-2\beta\Delta E}+6e^{-\beta\Delta E}+1\right)\Delta E^2}{4\Delta t\left(e^{-\beta \Delta E}+1\right)^2}.\label{turw}
\end{equation}
As the system returns, on average, to the same state after each period, its entropy remains unchanged, and  average entropy production per cycle is equal to average heat output over temperature:
\begin{equation}
\Delta_i S=\frac{\overline{j}\Delta t}{T}.
\end{equation}
The thermodynamic uncertainty relation can now be verified:
\begin{multline}
\frac{\overline{ j}^2\Delta t}{\overline{\delta j^2}}=\frac{\left(1-e^{-\beta\Delta E}\right)^2}{1+6e^{-\beta\Delta E}+e^{-2\beta\Delta E}}\\\leq \frac{\exp\left(\frac{\beta \Delta E}{2}\frac{e^{-\beta\Delta E}-1}{e^{-\beta\Delta E}+1}\right)-1}{2}= \frac{e^{\Delta_i S/{k_B}}-1}{2},
\end{multline}
cf.~Fig.~\ref{tur1}. Again, the bound is tight in the limit of small $\Delta E$. Furthermore, the continuous time uncertainty relation, Eq.~(\ref{turor}), can be violated for both examples.

\section{Discussion}
In this letter, we have derived a generalized thermodynamic uncertainty relation valid for Markov chains, and time-symmetric, periodically driven systems. 
Some remarks are in place. First, it should  be possible to test this bound experimentally  \cite{blickle2006thermodynamics,martinez2016brownian,proesmans2016brownian,PhysRevX.7.021051}.  Second, the discrete time setting is particularly interesting in the context of information processing, which naturally occurs via discrete steps. Third, we stress that  the ingredients of our derivation are of mathematical and statistical nature. It would  be of interest to investigate how genuine thermodynamic information allows to possibly refine the bounds and give them additional meaning.
Finally, the question remains whether an uncertainty relation can be derived for  systems with time-asymmetric driving. One might combine the results from this paper with the bound found in \cite{shiraishi2016universal}.

\acknowledgments
We thank Patrick Pietzonka and Grant Rotskoff for helpful conversations.

\section{Appendix: proof of Eq.~(\ref{sb})}
We first use Jensen's inequality to show that
\begin{eqnarray}
&&\ln\mathcal{N}+\frac{\Delta_i S}{k_B} = \ln\left(\sum_{k,l}\frac{p_{kl}p_{lk}}{p_{kl}+p_{lk}}\right)\nonumber\\
&&+\sum_{kl}\frac{p_{kl}-p_{lk}}{2}\ln\left(\frac{p_{kl}}{p_{lk}}\right)\nonumber\\& &\geq\sum_{k,l}\frac{(p_{kl}+p_{lk})}{2}\ln\left(\frac{2p_{kl}p_{lk}}{\left(p_{kl}+p_{lk}\right)^2}\right)\nonumber\\
&&+\sum_{k,l}\frac{p_{kl}-p_{lk}}{2}\ln \frac{p_{kl}}{p_{lk}}\\
&&=\sum_{k,l}p_{kl}\left(\frac{1+u_{kl}}{2}\ln\frac{2u_{kl}}{\left(1+u_{kl}\right)^2}+\frac{u_{kl}-1}{2}\ln u_{kl} \right),\nonumber
\end{eqnarray}
with $u_{kl}=p_{lk}/p_{kl}$. One verifies that:
\begin{multline}
\frac{1+u}{2}\ln\frac{2u}{\left(1+u\right)^2}+\frac{u-1}{2}\ln u\\\geq(1-\ln 2)\frac{1+u}{2}-\frac{2u}{u+1},\,\;\forall u> 0.
\end{multline}
Applying this to the previous inequality gives:
\begin{eqnarray}
\ln\mathcal{N}+\frac{\Delta_i S}{k_B}&\geq& (1-\ln 2)\sum_{k,l}\frac{p_{kl}+p_{lk}}{2}-2\sum_{k,l}\frac{p_{kl}p_{lk}}{p_{kl}+p_{lk}}\nonumber\\
&=&1-\ln 2-2\mathcal{N}.
\end{eqnarray}
With
$
1-\ln 2-2\mathcal{N}\geq \ln\left(1-\mathcal{N}\right), \forall \mathcal{N}\geq 0
$, one arrives at
\begin{equation}
\ln\mathcal{N}+\frac{\Delta_i S}{k_B}\geq \ln\left(1-\mathcal{N}\right),
\end{equation}
hence Eq.~(\ref{sb}).

\section{Appendix: large deviation function of empirical paths}
To derive the large deviation function of the empirical density, cf.~Eq.~(\ref{edldf2}), we first 
consider the probability distribution for $\{q_\Gamma\}$:
\begin{equation}
P_t(\{q_\Gamma\})=P_t(\{q_\Gamma\}|{\bf q})P_t({\bf q}),
\end{equation}
where ${\bf q}=\{q_{kl}\}$, $q_{kl}$ being the fraction of cycles which start at state $k$ and end at state $l$. The associated large deviation function is given by
\begin{equation}
I(\{q_\Gamma\})=-\lim_{t\rightarrow\infty}\frac{1}{t}\ln P_t(\{q_\Gamma\})=I(\{q_\Gamma\}|{\bf q})+I({\bf q}).
\end{equation}
Since the transition between initial and final states after each period is described by a Markov chain, $I({\bf q})$ is given by Eq.~(\ref{edldf}). Furthermore, consecutive cycles are independent, hence{, omitting some mathematical details concerning the summation of paths, one writes}:
\begin{equation}
P_t(\{q_\Gamma\}|{\bf q})=\prod_{k,l}\left( \frac{(Nq_{kl})!}{\displaystyle{\prod_{\{\Gamma_{kl}\}}} (Nq_\Gamma)!}\displaystyle{\prod_{\{\Gamma_{kl}\}}} \left(\frac{p_{\Gamma}}{p_{kl}}\right)^{Nq_{\Gamma}}\right),
\end{equation}
where $N=t/dt$ and $\{\Gamma_{ij}\} $ the set of trajectories starting in state $i$ and ending in state $j$. Using Stirling's approximation, one can now derive the expression for the conditional large deviation function, $I(\{q_\Gamma\}|{\bf q})$:
\begin{equation}
I(\{q_\Gamma\}|{\bf q})=\frac{1}{\Delta t}\left(\sum_{\Gamma} q_\Gamma\ln\left(\frac{q_{\Gamma}}{p_{\Gamma}}\right)-\sum_{k,l}q_{kl}\ln\left(\frac{q_{kl}}{p_{kl}}\right)\right)
\end{equation}
Combination with the large deviation function for $I({\bf q})$ leads to Eq.~(\ref{edldf2}). 


\end{document}